\documentclass[preprint,showpacs,preprintnumbers,amsmath,%
amssymb,superscriptaddress,floatfix]{revtex4}

\usepackage{graphicx}
\usepackage{dcolumn}
\usepackage{bm}
\newcommand{\be}{\begin{equation}}
\newcommand{\ee}{\end{equation}}
\newcommand{\bea}{\begin{eqnarray}}
\newcommand{\eea}{\end{eqnarray}}
\newcommand{\ba}{\begin{array}}
\newcommand{\ea}{\end{array}}

\begin{document}

\title{Domain wall dynamics in expanding spaces}

\author{Francisco J. Cao}
\email{francao@fis.ucm.es}
\affiliation{Departamento de F\'{\i}sica At\'omica, Molecular y
Nuclear, Universidad Complutense de Madrid, Avenida Complutense s/n,
E-28040 Madrid, Spain.} 
\affiliation{LERMA, Observatoire de Paris. Laboratoire
Associ\'e au CNRS UMR 8112, 61, Avenue de l'Observatoire, 75014
Paris, France.\\}

\author{El\'{\i}as Zamora-Sillero}
\email{elias@euler.us.es}
\affiliation{Departamento de F\'\i sica Aplicada I, E.\ U.\ P.,
Universidad de Sevilla, Virgen de \'Africa 7, 41011 Sevilla,
Spain}

\author{Niurka R.\ Quintero}
\email{niurka@euler.us.es}
\affiliation{Departamento de F\'\i sica Aplicada I, E.\ U.\ P.,
Universidad de Sevilla, Virgen de \'Africa 7, 41011 Sevilla,
Spain}

\date{\today}

\begin{abstract}
We study the effects on the dynamics of kinks due to expansions
 and contractions of the space. We show that the propagation
 velocity of the kink can be adiabatically tuned through slow
 expansions/contractions, while its width is given as a function of
 the velocity. 
We also analyze the case of fast expansions/contractions, where we are no
longer on the adiabatic regime. In this case the kink moves more slowly after
an expansion-contraction cycle as a consequence of loss of energy through
radiation. All these
 effects are numerically studied in the nonlinear Klein-Gordon 
equations (both for the sine-Gordon and for the $ \phi^4 $
 potential), and they are also studied within the framework of the
 collective coordinate evolution equations for the width and the
 center of mass of the kink. These collective coordinate evolution
 equations are obtained with a procedure that allows us to consider
 even the case of large expansions/contractions.
\end{abstract}


\maketitle


\section{Introduction}

Domain walls are the interface between two regions where the
system is close to two different stable or metastable states.
Their formation and dynamics are relevant for many domains of
Physics as Solid State Physics (magnetic materials)
\cite{scottbook,prb19}, Liquid Crystals Science \cite{kalman},
Plasma Physics \cite{prl88,pp11} and Cosmology (phase transitions
in the Early Universe) \cite{linde,oliveira}.
Spatially extended systems with more than one stable or
metastable state can present domain walls, as is the case of
systems that have undergone a first order phase transition.
An expansion or contraction of these systems can be realized through
an external action on the
system (magnetic materials, liquid crystals, plasmas), or they are
already present in the system dynamics (phase transitions in the
Early Universe).

The kink solutions of the one-dimensional nonlinear Klein Gordon
equation for potentials with two or more stable states (as
sine-Gordon or $ \varphi^4 $) provide models for the propagation
of plane domain walls in the direction perpendicular to the
domain wall. These models have been successfully used in several
contexts as in the description of the dynamics of domain walls in
ferrodistortive materials \cite{prb19}, and the
magnetohydrodynamic mode trajectory in reversed-field pinch
experiments \cite{prl88,pp11}. On the other hand, expansions and
contractions can be realized in magnetic materials through the
mechanical action on the material \cite{defmat}, and in plasmas
through confinement \cite{confplasma}; while in the case of phase
transitions in the Early Universe, the expansion is already
present in the system dynamics. These expansions/contractions can
be parameterized with a scale factor $ a(t) $ when they are
homogeneous in space \cite{weinberg}. The characteristic times for the time
variation of $ a(t) $ can be shorter or longer than the
characteristic times of the kink dynamic.

With these facts in mind, in this paper we consider kinks of the nonlinear
Klein-Gordon equations for the sine-Gordon and the $ \phi^4 $ potentials
(as a paradigmatic examples of integrable and non integrable nonlinear
Klein-Gordon equations, respectively), and we study theirs dynamics in slow and fast 
expanding/contracting spaces.

In order to achieve this purpose, our paper is organized as
follows: In Section \ref{sKGexpanding} we introduce the nonlinear
Klein-Gordon equations in expanding/contracting spaces, and derive
the continuity equations for the energy and momentum densities. In
the next section, Section \ref{sLagrangian}, we use the Rice {\it
Ansatz} \cite{rice} in these continuity equations, and we reduce
the problem with an infinite number of degrees of freedom to an
approximate description in terms of two collective coordinates
(CC), the center and the width of the kink. In Section
\ref{sKinkDyn}, we study the kink dynamic in both slow and fast
expanding/contracting spaces using the collective coordinate
evolution equations, and integrating numerically the complete
nonlinear Klein-Gordon equations. We first consider the case of
slow expansion/contraction within the adiabatic approximation, and
later we discuss the fast expansion/contraction case and the
departures from the adiabaticity that emerge. Finally, in the
conclusions, we summarize and discuss the results.

The previous sections are complemented with two appendices where some
technical details related with expanding/contracting spaces are
reviewed, and the equivalence between several collective coordinates
approaches \cite{gtwa,emderivation,lagform} to this problem is shown.

\section{Klein-Gordon equation in expanding spaces}\label{sKGexpanding}
The Klein-Gordon equation in non expanding spaces can be obtained
by extremizing the action defined by the Lagrangian density,

\be S = \int{dt\,dx\,{\cal L}} = \int{dt\,dx\, \left[ \frac12
\phi_t^2 - \frac12 \phi_x^2 - U(\phi) \right] } \;. \ee

In an expanding (or contracting) space with one spatial dimension
the space-time distance between events is given by \be ds^2 = dt^2
- a^2(t) dx^2 \;, \ee where $ a(t) $ is the scale factor that
gives the dilation of the physical spatial distances (see Appendix A). 
Thus, the
physical spatial distance between two points with coordinates $
x_1 $ and $ x_2 $ is
\be
d_{12}(t) = a(t) |x_2-x_1| \;. \label{eq3}
\ee

The Klein-Gordon equation in this expanding space can be obtained
noting that after the spatial dilation $ dx \to a(t) dx $ the
action for a Klein-Gordon field is
\be
S = \int{dt\,dx\,a(t){\cal L}} = \int{dt\,dx\,a(t) \left[ \frac12
\phi_t^2 - \frac12 \frac{\phi_x^2}{a^2(t)} - U(\phi) \right] } \;,
\ee
where the subscripts $ t $ and $ x $ indicate the time and space
partial derivatives respectively; and $ U(\phi) $ is the nonlinear
Klein-Gordon potential.

The Euler-Lagrange equation for the field $ \phi $,
\be
\frac{\partial\;}{\partial t} \frac{ \delta{[a(t) \cal
L]}}{\delta\phi_t} + \frac{\partial\;}{\partial x} \frac{
\delta{[a(t) \cal L]}}{\delta\phi_x} - \frac{ \delta{[a(t) \cal
L]}}{\delta\phi} = 0 \;,
\ee
gives the equation of motion,
\be \label{KGevoleq}
\phi_{tt} + H(t) \phi_t - \frac{\phi_{xx}}{a^2(t)} +
\frac{dU}{d\phi} = 0 \;,
\ee
with $ H = a_t/a $.

We have to keep in mind that the physical distance is $
\displaystyle a(t)|x_2-x_1|$ and not $ |x_2-x_1|$ (the latter is
usually called the comoving distance). Thus, when $\displaystyle
a(t)$ grows the space experiences an elongation and when
$\displaystyle a(t)$ decreases the space experiences a contraction.

We define the energy density $ \rho_E(t,x) $, the momentum density
$ \rho_P(t,x) $, and the momentum current $j_P(t,x) $, as the
time-time, space-time and space-space contravariant components,
respectively, of the energy-momentum tensor. We obtain
\bea
 \rho_E &=& \frac12 \phi_t^2 + \frac12 \frac{\phi_x^2}{a^2} +
U(\phi)\;, \label{rhoE} \\
 \rho_P &=& -\frac{1}{a^2} \phi_t \phi_x\;, \label{rhoP} \\
 j_P &=& \frac{1}{a^2} \left( \frac12 \phi_t^2 + \frac12 \frac{\phi_x^2}{a^2} -
 U(\phi) \right) \;. \label{jP}
\eea
The energy-momentum conservation (in the covariant sense) gives
the following relations
\bea \label{rhoEconserv}
&& \frac{\partial \rho_E}{\partial t} + H (\rho_E + a^2 j_P) +
\frac{\partial \rho_P}{\partial x} = 0 \;,\\
&& \frac{\partial \rho_P}{\partial t} + 3 H \rho_P +
\frac{\partial j_P}{\partial x} =0 \;, \label{rhoPconserv}
\eea
that also can be obtained using the equation of motion
\eqref{KGevoleq}.

This also implies relations between the total energy $ E =
\int{a\,dx \; \rho_E} \;,$ the total momentum $ P = \int{a\,dx \;
\rho_P} \;, $ and the total momentum current $ J_P = \int{a\,dx \;
j_P} \;. $ They are given by 
\be \label{uno}
\frac{d E}{dt} + H a^2 J_P = 0,
\ee
[provided $ \rho_P(t,\infty) = \rho_P(t,-\infty) $],  and
\be \label{dos}
 \frac{d P}{dt} + 2 H P = 0,
\ee
[provided $ j_P (t,\infty) = j_P(t,-\infty) $]. Note that for a
space that is neither expanding nor contracting [i.e., $a(t)=1$] $E$ and $P$ are
conserved quantities \cite{scottbook}.

\section{Collective Coordinates analysis for perturbed Klein-Gordon
equation}\label{sLagrangian}

In the previous section we have shown that
the momentum and the energy are no longer
conserved quantities in expanding/contracting spaces, instead they
satisfy the continuity equations (\ref{uno}) and (\ref{dos}),
respectively. We are interested in the evolution of the kink that we
describe through two collective coordinates, its width and its center of
mass position. Therefore, we neglect the possible energy transfer to
other degrees of freedom.

In order to obtain the equations of motion for these two
collective coordinates (CCs) we use the Rice {\it Ansatz}
\cite{rice} in the continuity equations (\ref{uno}) and
(\ref{dos}) [with the energy and momentum given by
Eqs.~\eqref{rhoE}-\eqref{jP}], and we study two particular
nonlinear potentials, the sine-Gordon (sG) potential
$U(\phi)=1-\cos(\phi)$, and the $\phi^4$ potential
$U(\phi)= \frac{1}{4}\,(1-\phi^2)^2$. The domain walls in these systems
are represented by kink-like solutions that describe the
transition between two regions in different minima of the
potential. For the sine-Gordon potential the Rice {\it Ansatz}
assumes the function
\be \label{rice-sg}
\phi(x,t) = 4 \, \arctan \left\{ \exp
\left[\frac{x-X(t)}{l(t)} \right] \right\},
\ee
whereas for the $ \phi^4 $, the Rice {\it Ansatz} reads
\be \label{rice-phi4}
\phi(x,t) = \tanh \left[\frac{x-X(t)}{l(t)}\right],
\ee
where $X(t)$ and $l(t)$ represent in both cases the center
and the width of the kink, respectively, in the comoving frame.
Note that for the unperturbed system [take in Eq.~(\ref{KGevoleq})
$a(t)=1$] the kink-like
solutions are represented by Eqs.\ (\ref{rice-sg}) and
(\ref{rice-phi4}) with $X(t)=v t$ and $l(t)=l_{s}=l_{0}
\sqrt{1-v^2}$, being $v$ and $l_{s}$ the constant velocity and
width of the kink, respectively; and $l_{0}=1$ ($l_{0}=\sqrt{2}$)
the width of the soliton at rest for sG ($\varphi^4$). The time
dependence of the scale factor $a(t)$ (expansions/contractions)
implies changes in the width of the kink, as we show later,
justifying the choice of the Rice {\it Ansatz}.

Inserting the {\it Ansatz}, Eq.~(\ref{rice-sg})
or (\ref{rice-phi4}), in the expressions for $\rho_{P}$, $\rho_E$,
and $j_{P}$ and integrating over $x$, we obtain
\bea
P(t)&=& \frac{M_{0} l_{0} \dot{X}}{a(t)\,l}, \\
\label{energy}
E(t)&=&\frac{M_{0}\,l_{0}}{2}\frac{a\dot{X}^{2}}{l}
+\frac{\alpha\,M_{0}\,l_{0}}{2}
\frac{a\,\dot{l}^{2}}{l}+\frac{M_{0}}{2}\left(\frac{l_{0}}{a\,l}+
\frac{a\,l}{l_{0}}\right),\\ \label{fluxP}
J_{P}(t)&=&\frac{M_{0}\,l_{0}}{2}\frac{\dot{X}^{2}}{a\,l}
+\frac{\alpha\,M_{0}\,l_{0}}{2}
\frac{\dot{l}^{2}}{a\,l}+\frac{M_{0}}{2\,a^{2}}\left(\frac{l_{0}}{a\,l}-
\frac{a\,l}{l_{0}}\right),
\eea
respectively; where $\alpha=\pi^2/12$ and $M_0=8$
for the sine-Gordon and $\alpha=(\pi^2-6)/12$ and $M_0=2\sqrt{2}/3$
for $\varphi^4$. These expressions are replaced in the continuity
equations (\ref{uno}) and (\ref{dos}), giving the following systems of
ordinary differential equations (ODE)
for $X(t)$ and $l(t)$:
\begin{eqnarray} \label{ccex}
&& \dot{X} = \frac{a(t) P(t) l(t)}{M_{0} l_{0}}, \\ \label{ccep}
&& \frac{dP}{dt} =- 2H  P, \\
\label{ccel}
&& \alpha \left [ \dot{l}^2-2 H l\dot{l} -2l\ddot{l}\right] = \frac{l^2}{l^2_0} \left
(1+a^2(t)\frac{P^2}{ M_0^2} \right )-\frac{1}{a^2},
\end{eqnarray}
where the dots denote the derivative with respect to $t$. If the
initial conditions are those of a kink with initial position $
X(0) $ and initial velocity 
$ \dot X(0) $, this implies $ l(0) = l_0 \sqrt{1-\dot
X^2(0)} $ and $ \dot l(0) = 0 $, and $ P(0) $ is determined by
Eq.~\eqref{ccex} at $ t=0 $. The same collective coordinates
evolution equations are obtained using the Lagrangian method
\cite{lagform,saan} or the Generalized Travelling Wave Ansatz
(GTWA) \cite{gtwa,prl} (based on projection techniques), as it is
shown in Appendix \ref{equivalencia}.

This set of evolution equations, Eqs.~\eqref{ccex}-\eqref{ccel}, involves the
variables $X(t)$ and $l(t)$ through the momentum $P(t)$ defined in Section
\ref{sKGexpanding}. Note that the equation
for the momentum is linear, and therefore it can be solved exactly yielding
\be \label{ccp}
P(t)= \frac{P(0)}{a^{2}(t)},
\ee
with $P(0)=\dot{X}(0) M_{0} l_{0}/\sqrt{1-\dot{X}^{2}(0)}$. The other 
two equations are nonlinear and coupled, and they are analyzed in detail
in the next section.

\bigskip

\section{Kink dynamics in expanding/contracting
spaces}\label{sKinkDyn}

First of all, let us remark that the main physical variables associated
with the kink propagation are the physical center of the kink,
$X_{phys}(t)=\displaystyle a(t)X(t)$, its physical width
$l_{phys}(t)=\displaystyle a(t)l(t)$, and its physical velocity
\begin{equation} \label{vkink}
V_{phys}(t)=H X_{phys}(t)+a(t)\frac{d X(t)}{d t} \;.
\end{equation}
The first term on the r.h.s. is the contribution to the
kink velocity due to the expansion or contraction of the space
(i.e., it turns to zero when the space finishes its elongation or
contraction $ H=0 $). On the other hand, the second term is the peculiar
velocity of the kink with respect to the propagating space,
\be
V_{pec}(t)=a(t)\, \frac{d X(t)}{d t} \;.
\ee

For the sake of clarity, we analyze separately the cases of slow and fast
expansion/contraction.

\subsection{Slow expansion/contraction (adiabatic regime)}

Note from Eqs.\ (\ref{ccex}), (\ref{ccep}) and (\ref{ccel})
that the adiabatic approximation implies
\be \label{adiabcond}
\alpha \left [ \dot{l}^2-2 H l\dot{l} -2l\ddot{l}\right] \ll
\frac{1}{a^2} \;.
\ee
Hence, from Eq.~\eqref{ccel} we obtain the relation
\begin{equation} \label{ladia}
l_{phys}(t)= l_{0} \, \sqrt{1-V_{pec}^{2}(t)}\;,
\end{equation}
that links the physical
width and the peculiar velocity of the kink. Furthermore, from
Eqs.\ (\ref{ccex}) and (\ref{ccp})-(\ref{ladia}) we obtain
\begin{equation} \label{vadia}
V_{pec}(t)= \frac{P(0)}{\sqrt{M_{0}^{2} a^2(t) +P^2(0)}}.
\end{equation}
From the previous Eqs.~(\ref{ladia}) and (\ref{vadia}) we see that
an expansion decelerates the peculiar motion of the kink and makes
it wider, asymptotically
\be\label{lvfisExp}
\lim_{a(t)\rightarrow \infty} V_{pec}(t) = 0 \;,\quad
\lim_{a(t)\rightarrow \infty} l_{phys}(t) =  l_{0}.
\ee
On the other hand, a contraction accelerates the peculiar motion
of the kink and makes it sharper, asymptotically
\be\label{lvfisCont}
\lim_{a(t)\rightarrow 0} V_{pec}(t) = 1 \;,\quad
\lim_{a(t)\rightarrow 0} l_{phys}(t) = 0.
\ee

The adiabaticity conditions can be obtained from
Eq.~(\ref{adiabcond}) and require only slow expansions/contractions
\bea
|H| &\ll& \frac{\Omega_{R}}{1-V_{pec}^2} \;, \label{adiabH} \\
|\dot H| &\ll& \frac{\Omega_{R}^2}{(1-V_{pec}^2)^2} \;,
\label{adiabdotH}
\eea
where $\Omega_{R}=1/(\sqrt{\alpha} l_{0})$ is the so called Rice
frequency for zero velocity, 
$\Omega_{R}=\sqrt{12}/\pi=1.10... $ for sG and
  $\Omega_{R}=\sqrt{6/(\pi^2-6)}=1.24... $ for $\varphi^4$.

\begin{figure}
\begin{center}
\begin{tabular}{cc}
\includegraphics[width=5.5cm,angle=-90]{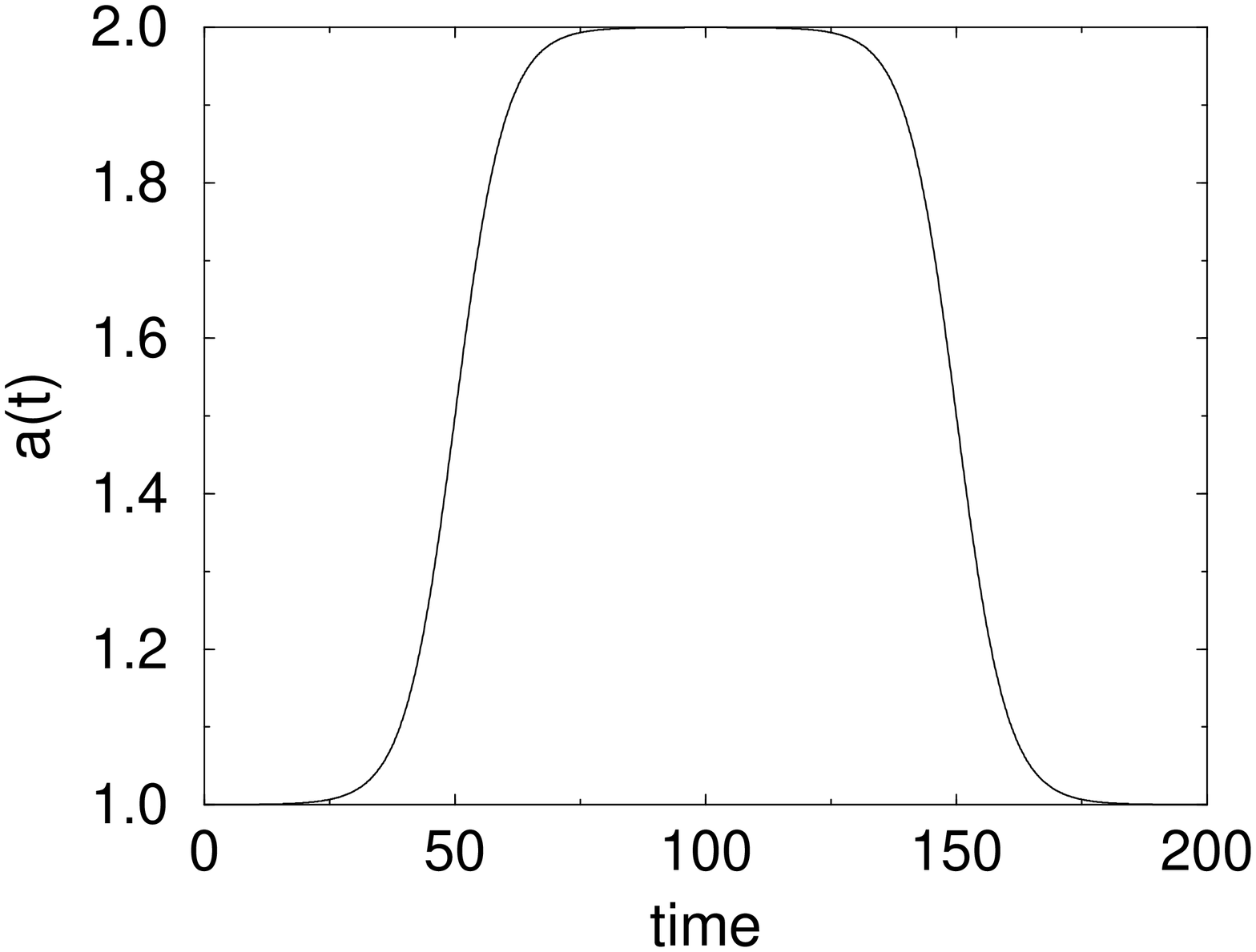}&
\includegraphics[width=5.5cm,angle=-90]{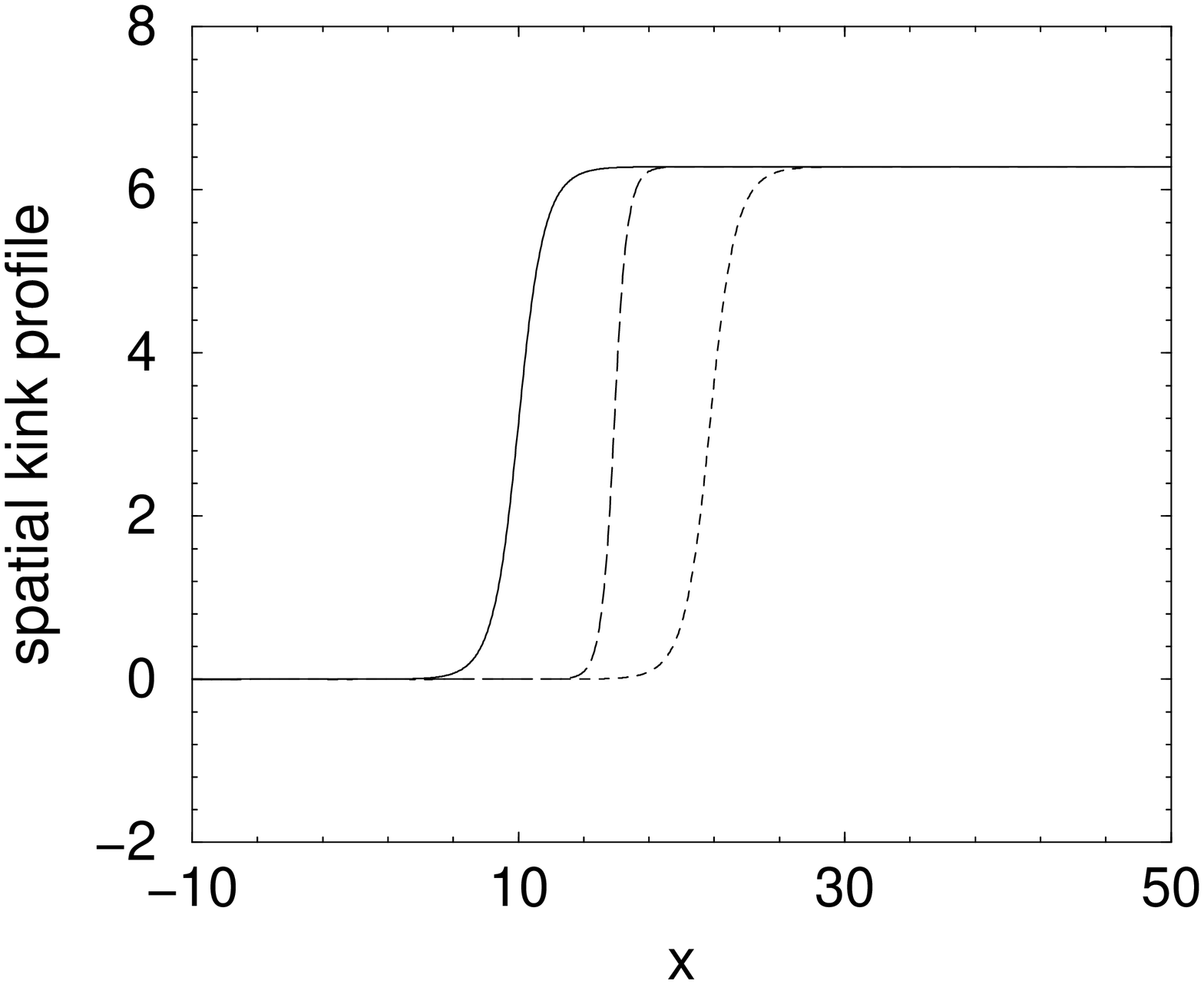}\\
\includegraphics[width=5.5cm,angle=-90]{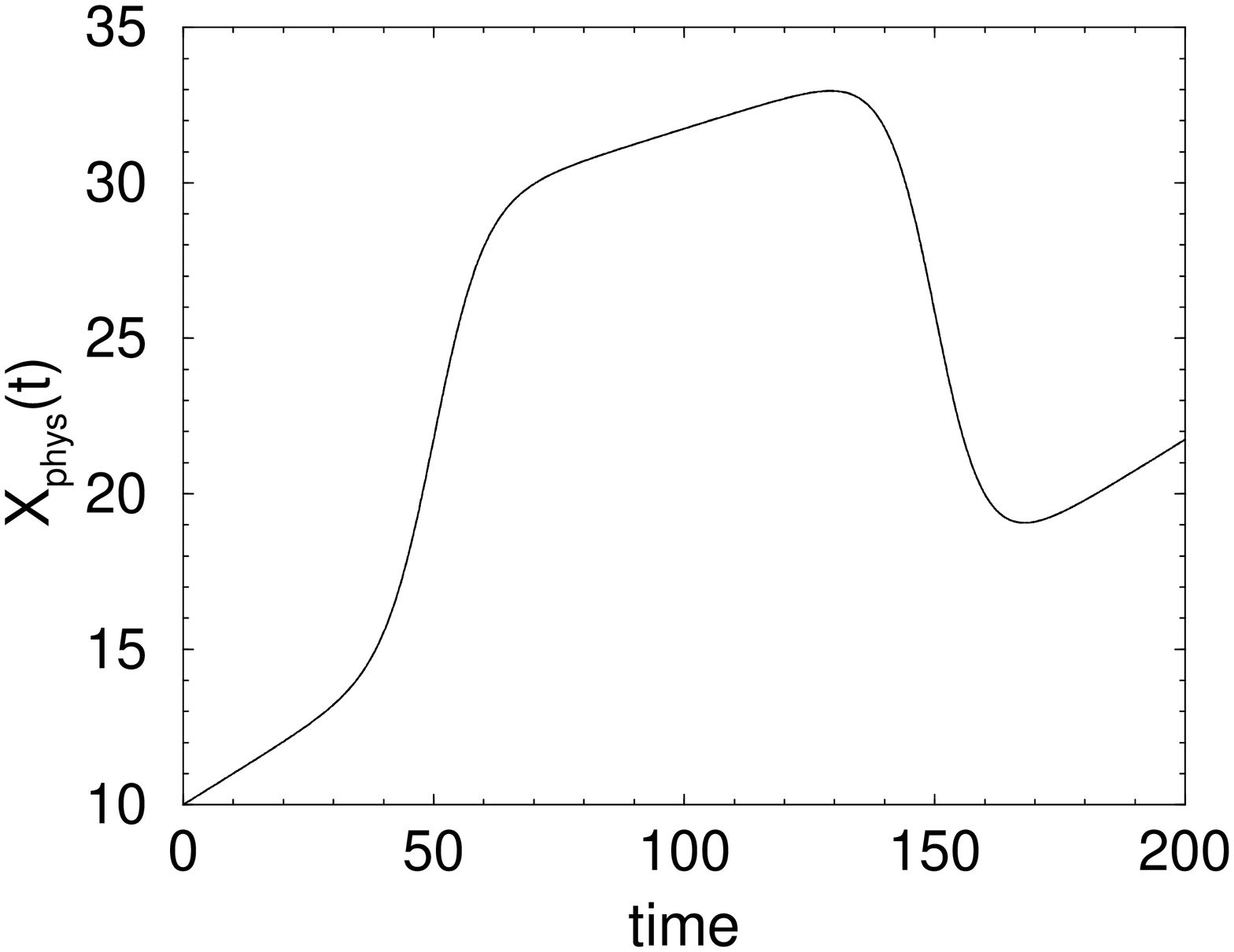}&
\includegraphics[width=5.5cm,angle=-90]{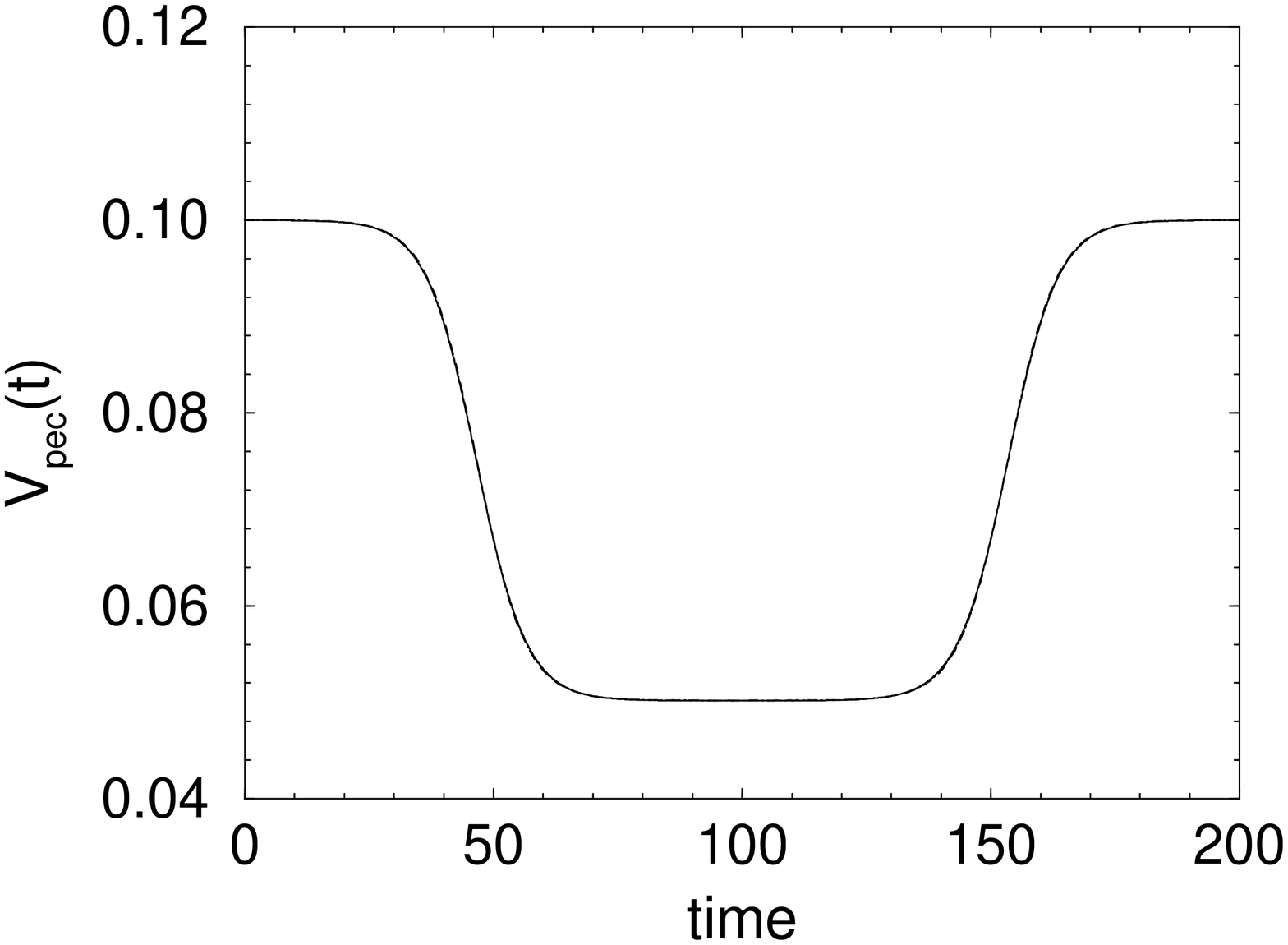}\\
\includegraphics[width=5.5cm,angle=-90]{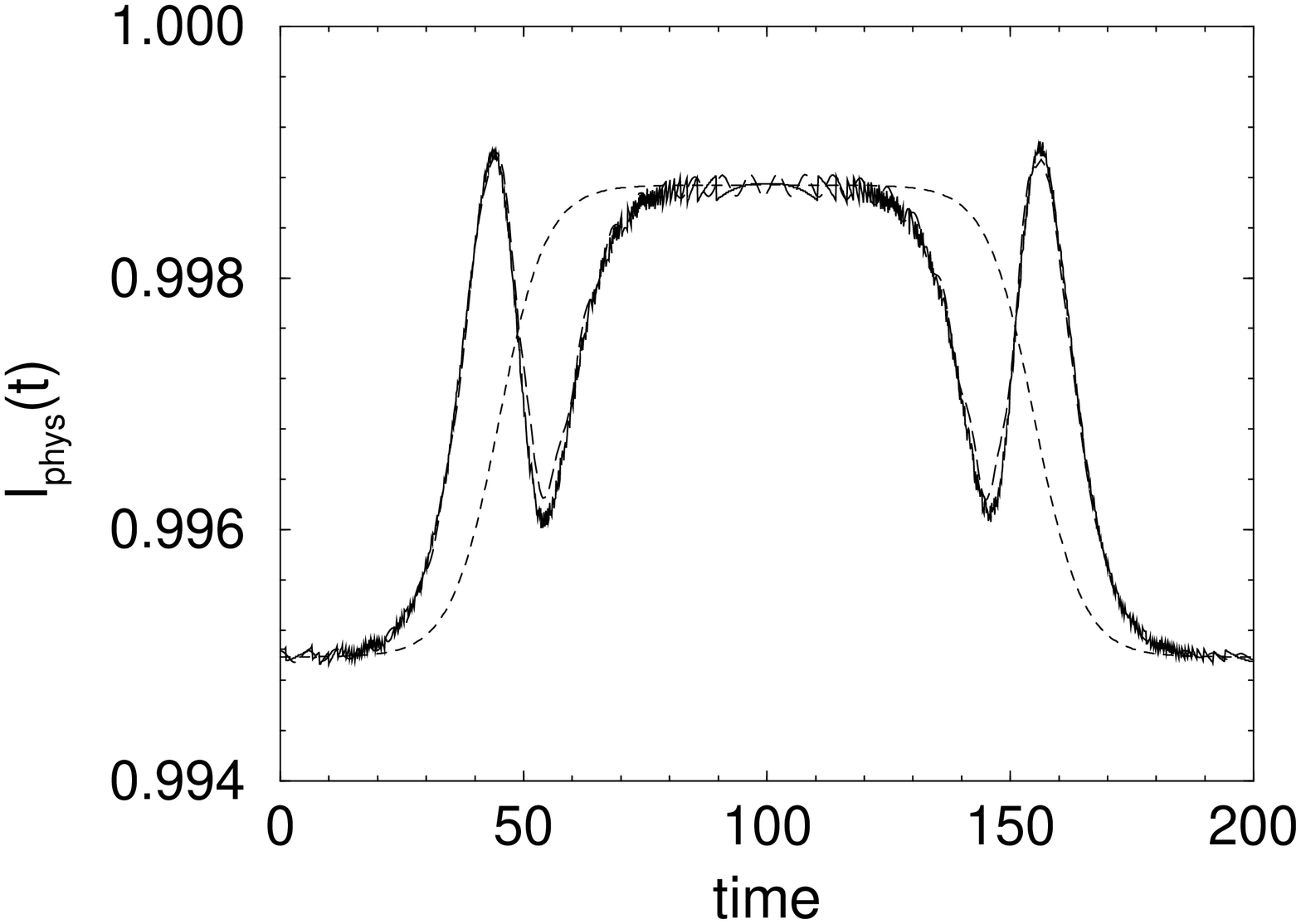}&
\includegraphics[width=5.5cm,angle=-90]{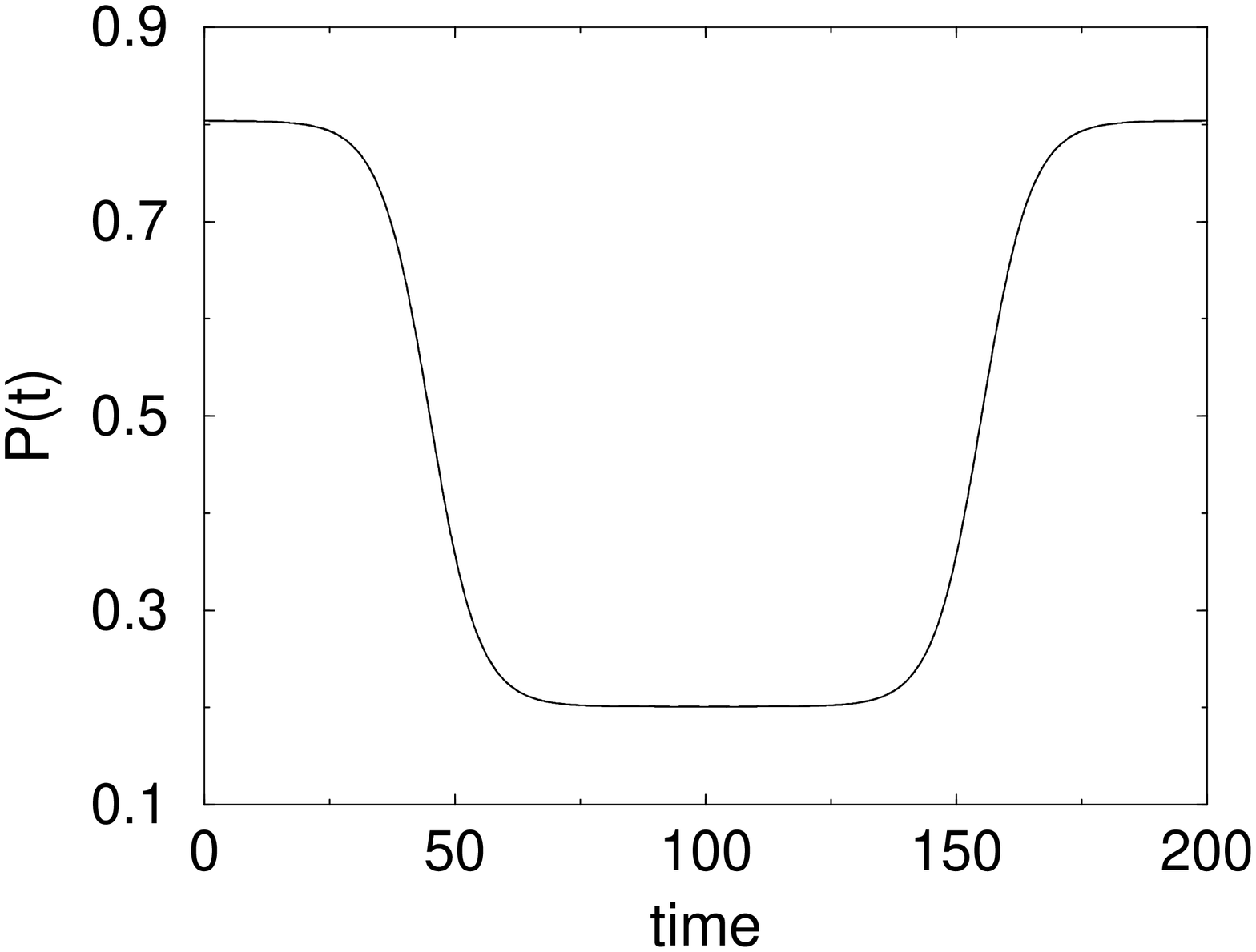}
\end{tabular}
\end{center}
\caption{Sine-Gordon kink dynamics in a \textit{slow} expanding
and later contracting space. Top-left panel: scale factor $ a(t)
$, Eq.~\eqref{adep} with $\Delta a = 1 $, $ t_0 = 50$, $ t_1 = 150
$ and $ \Delta t = 10 $. Top-right panel: spatial profile of the
kink $ \phi(x,t_{fix}) $ at times $t_{fix}=0$ (solid line),
$t_{fix}=100$ (long-dashed line) and $t_{fix}=200$ (dashed line).
Three following panels: position of the center of the kink $
X_{phys}(t) $, peculiar kink velocity $ V_{pec}(t) $, and width of
the kink $ l_{phys}(t) $ as functions of time given by the numerical 
simulations of the PDE
(solid line), the numerical solutions of the 
CCs equation (long-dashed line), and the
adiabatic approximation (dashed line). Bottom-right panel:
momentum of the kink $ P(t) $ exact solution. Initial conditions:
a kink with $ X(0) = 10 $ and $ \dot X(0) = 0.1 $. }
\label{figSlowExpCon}
\end{figure}

We have studied the dynamics for different expansion and
contraction rates, both for the sine-Gordon and the $ \phi^4 $
potentials. In particular, Fig.~\ref{figSlowExpCon} (and also
Fig.~\ref{figFastExpCon}) shows the results for an expansion followed
by a contraction parameterized by the function
\be \label{adep}
a(t)=1+\frac{\Delta a}{2}\, {\rm tanh\left(\frac{t-t_{0}}{\Delta t
}\right)} - \frac{\Delta a}{2}\, {\rm
tanh\left(\frac{t-t_{1}}{\Delta t}\right)},
\ee
with $\Delta a$, $t_{0}$, $t_{1}$ and $ \Delta t $ constants.
($\Delta a$ represents the changes in $a$; $t_{0}$ and $t_{1}$, the
times when the
expansion and contraction take place, respectively and $ \Delta t $ is the
characteristic time interval where the changes in $ a $ takes
place.)

In Fig.\ \ref{figSlowExpCon} we show the effects on a kink of a
\emph{slow} expansion followed by a slow contraction, that verify the
adiabatic conditions ($ H \sim \Delta a / \Delta t = 0.1 $ and $ \dot
H \sim \Delta a / (\Delta t)^2 = 0.01 $).
In this case, when the adiabatic conditions are verified, both
the CCs equations and the adiabatic approximations are in good
agreement with the exact results, especially for the center of the kink
$ X_{phys}(t) $ and its peculiar velocity $ V_{pec}(t) $ (see Fig.\
\ref{figSlowExpCon}). It is important to note that the agreement is good
even when the variations of $ a $ are large.

This implies that the main effects of an slow expansion/contraction are the
change of the width and the speed of propagation of the kink,
following the \emph{adiabatic relations} \eqref{ladia} and
\eqref{vadia}. Therefore, these results show that \emph{the speed of a kink can
be tuned} by slowly expanding/contracting the space.

Note that the CCs evolution equations go further and are able to
predict the small oscillations of the kink width \footnote{The
width and the center of the kink position for the PDE results are
obtained fitting the Rice \emph{Ansatz}, Eq.~\eqref{rice-sg} or
Eq.~\eqref{rice-phi4}, to the kink profile given by the PDE
evolution \eqref{KGevoleq}.} produced by the slow
expansion/contraction (see Fig.\ \ref{figSlowExpCon}). Therefore,
they can compute deviations from adiabaticity excluding
the radiative effects (because they involve the transfer of energy to other
degrees of freedom).

It is important to stress that the collective coordinate evolution
equations obtained with an \textit{Ansatz} with fixed width and
variable center of the kink does not predict any variation of the
kink speed due to expansions/contractions. Therefore, an
\textit{Ansatz} that allows the width of the kink to evolve, like 
the Rice \textit{Ansatz}, is an essential
ingredient in order to obtain the correct variation of the speed of the
kink under expansions/contractions. It is interesting to remark that, 
in the $\phi^4$ model the oscillations of the width are related  
to excitations of the internal mode \cite{rice}, 
while in the sine-Gordon equation it has been shown that the
excitation of certain phonons can imply oscillations in the shape of the kink
\cite{campbell,panos}.

On the other hand, in general 
the applicability of the perturbative approaches is restricted to 
slow and small expansions/contractions. The perturbative
approaches rewrite Eq. (\ref{KGevoleq}) in the form $ \phi_{tt} -
\phi_{xx} + dU/d\phi = \epsilon f \equiv -H(t) \phi_t - (1/a^2(t) - 1)
\phi_{xx} $, where $ \epsilon f $ is treated as a small perturbation.
Therefore, their range of applicability is limited not only 
to small $ H $ but also to 
small $(1/a^2 - 1)$, i.e., slow and small expansions/contractions.

\subsection{Fast expansion/contraction (non-adiabatic regime)}

\begin{figure}
\begin{center}
\begin{tabular}{cc}
\includegraphics[width=5.5cm,angle=-90]{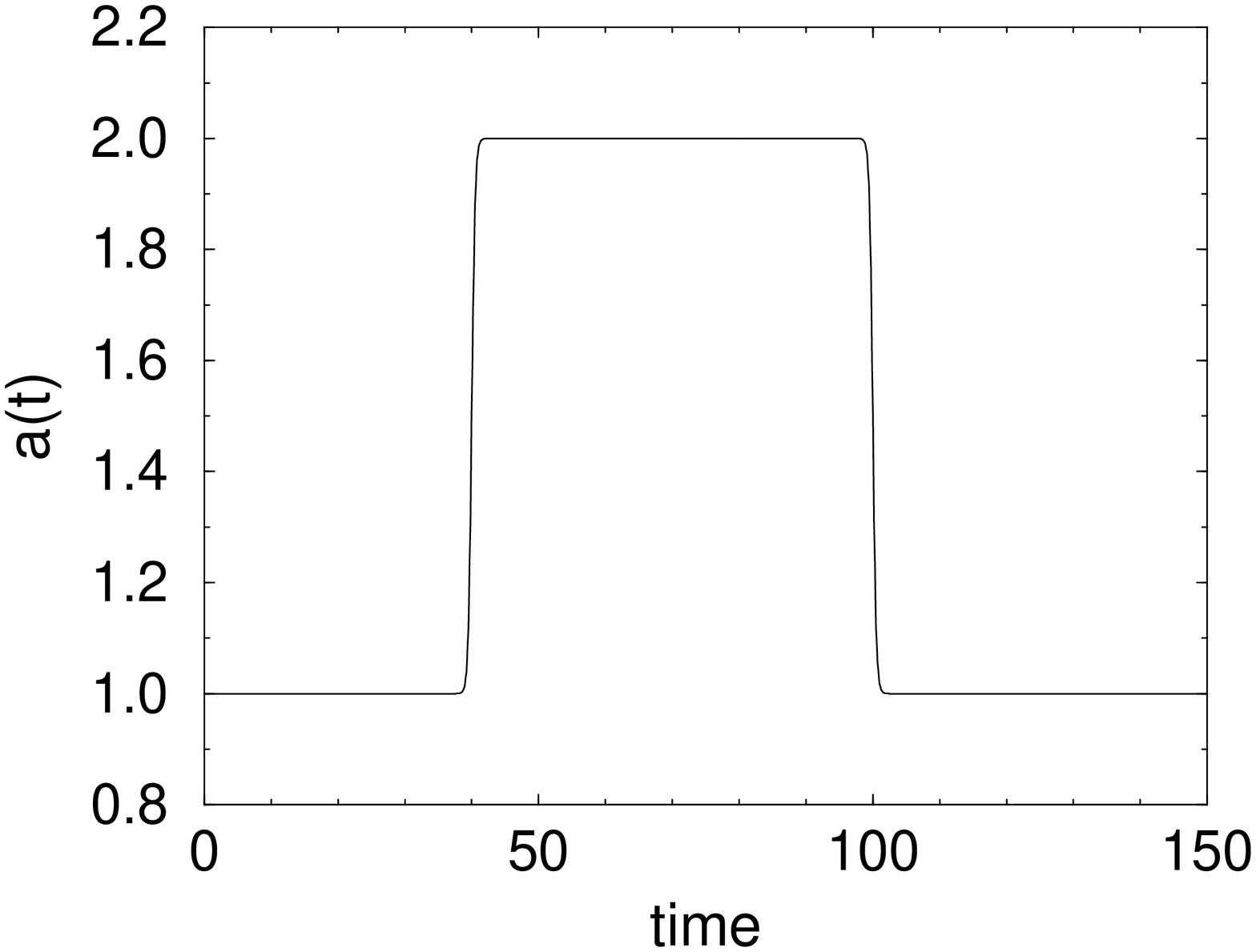}&
\includegraphics[width=5.5cm,angle=-90]{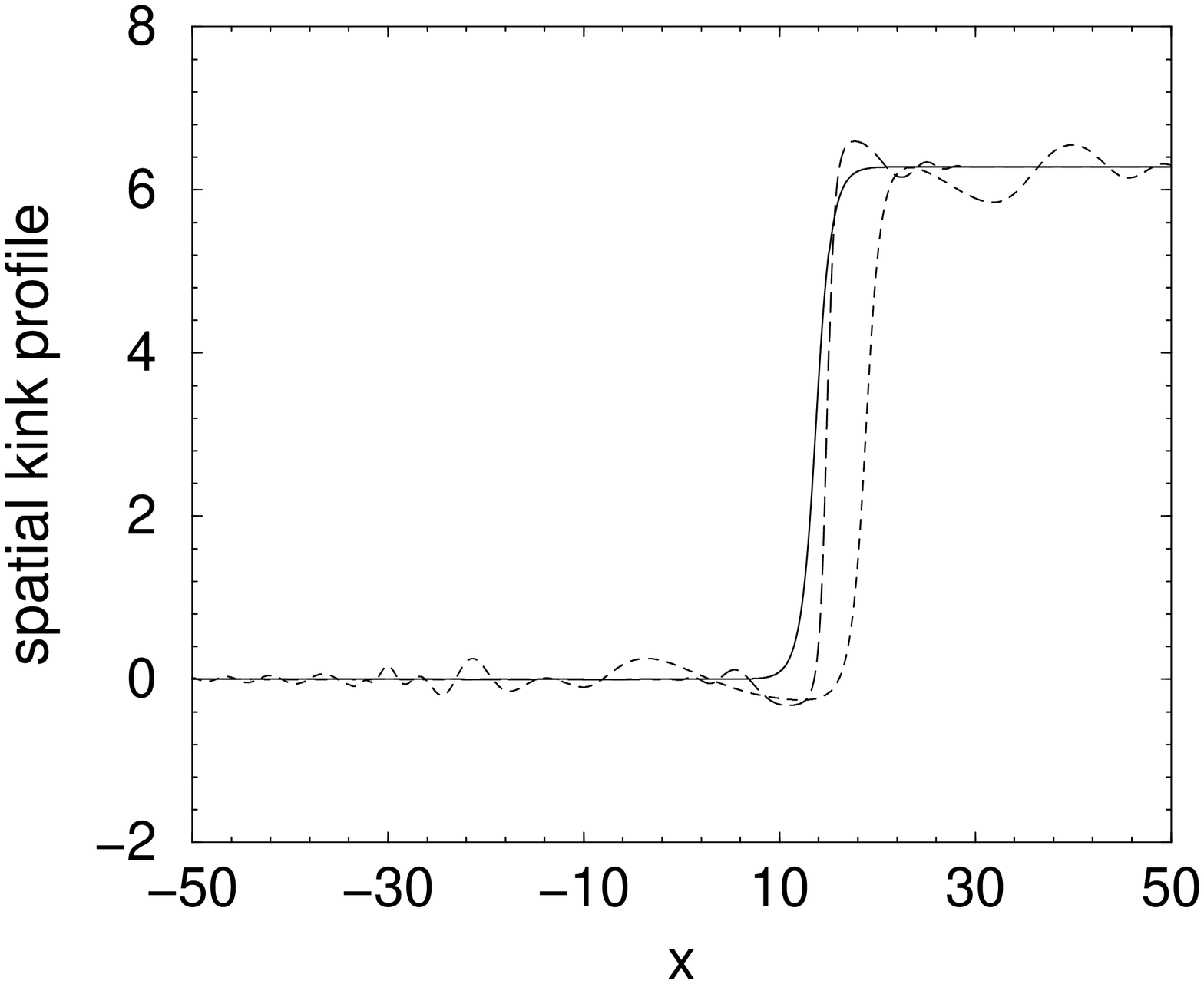}\\
\includegraphics[width=5.5cm,angle=-90]{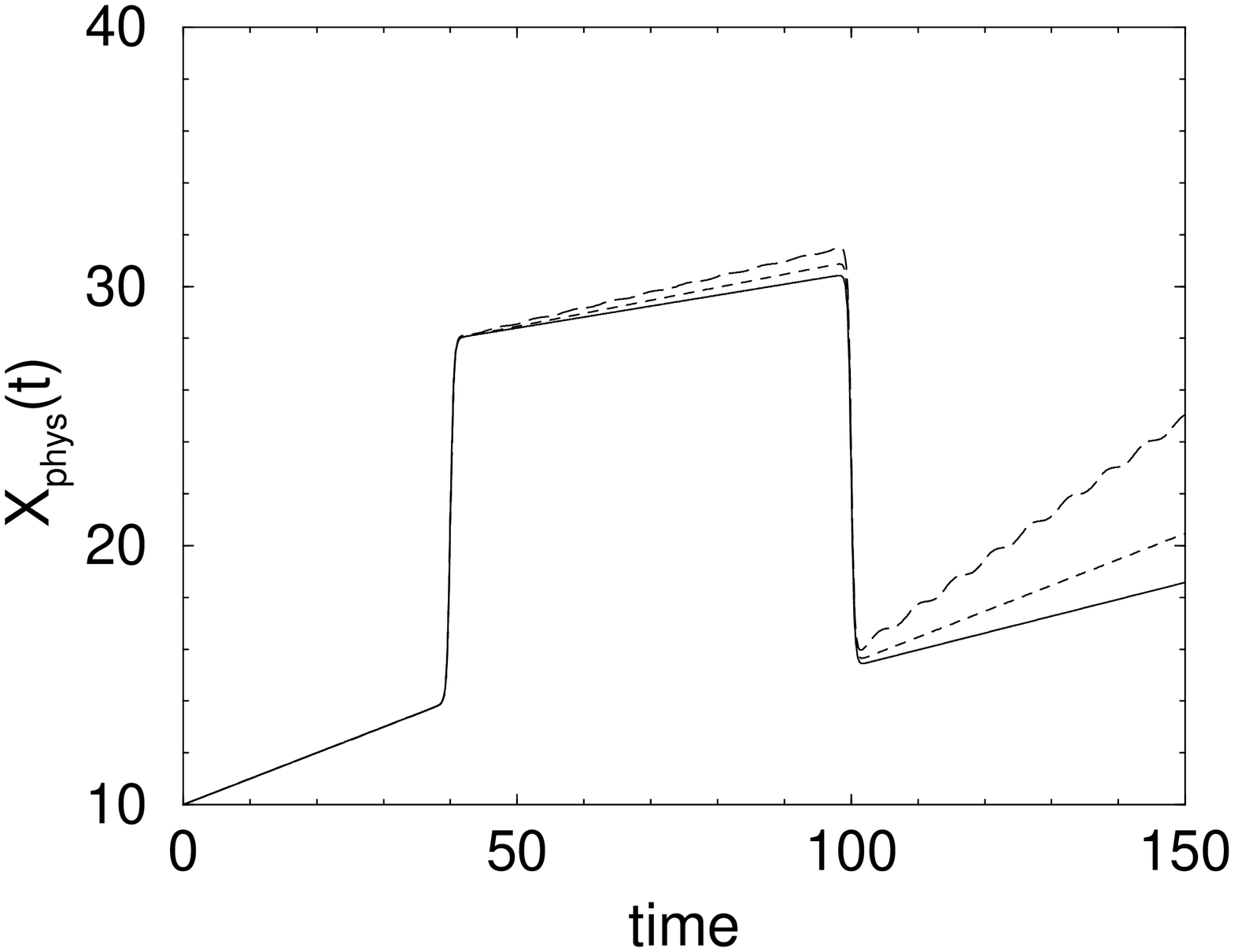}&
\includegraphics[width=5.5cm,angle=-90]{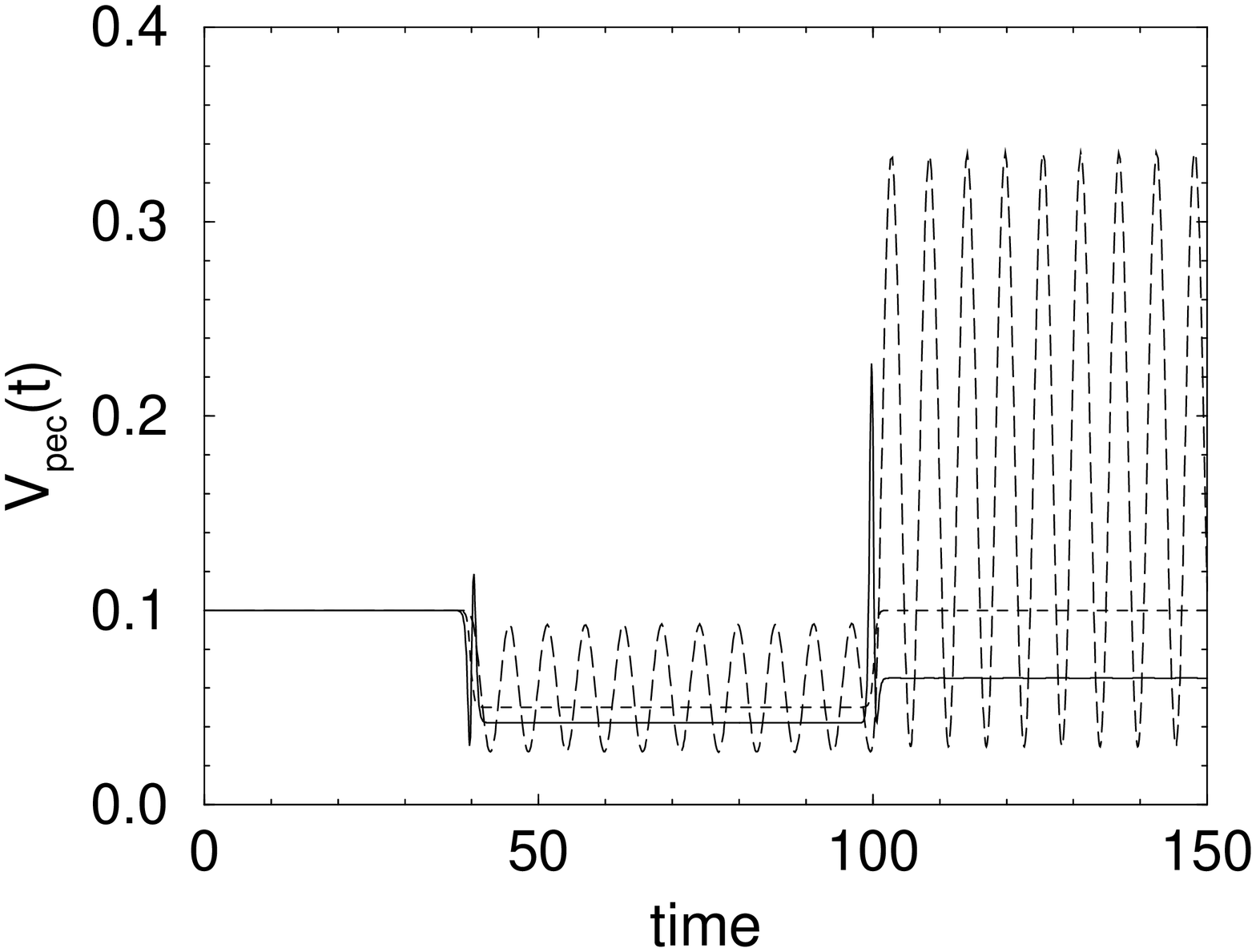}\\
\includegraphics[width=5.5cm,angle=-90]{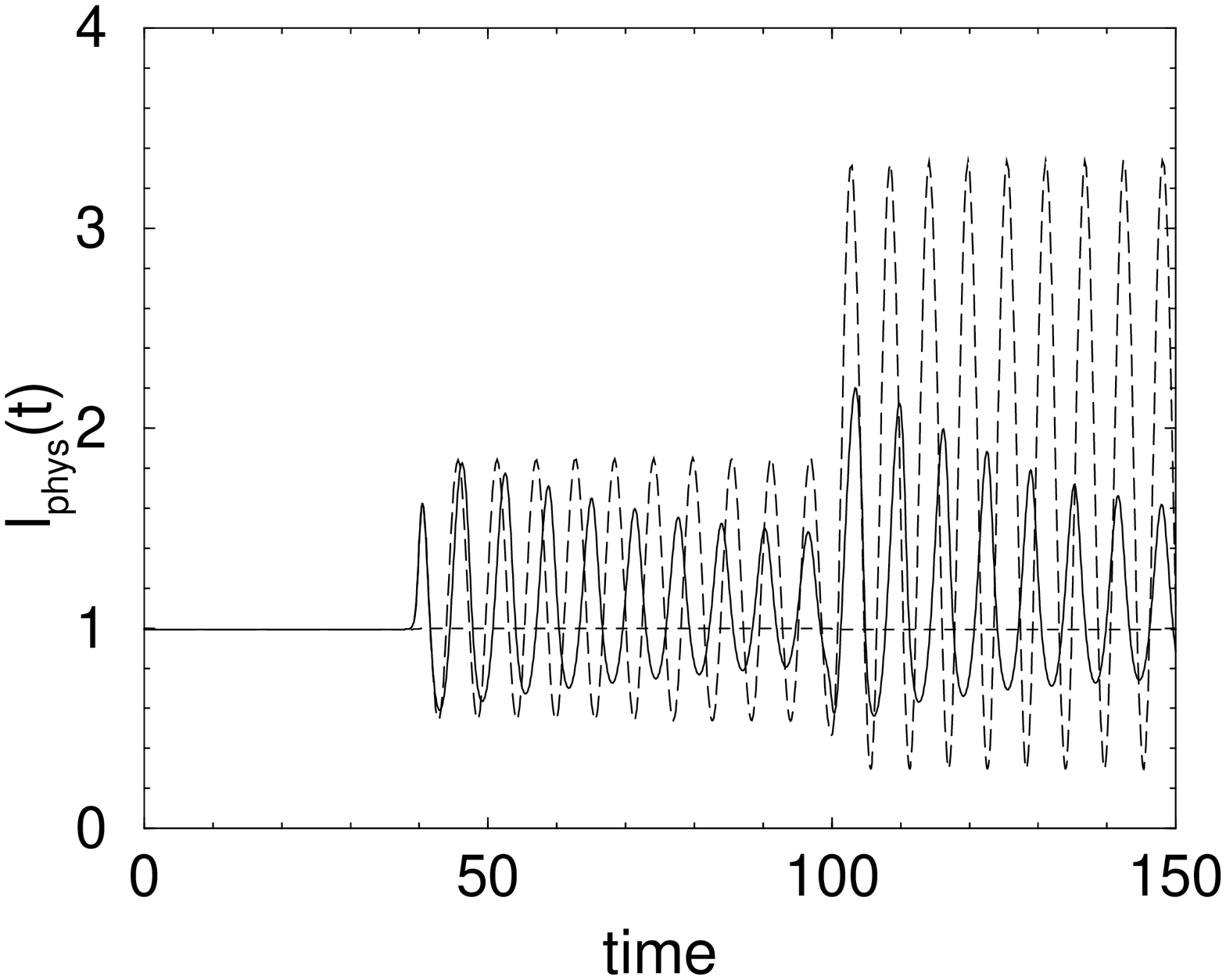}&
\includegraphics[width=5.5cm,angle=-90]{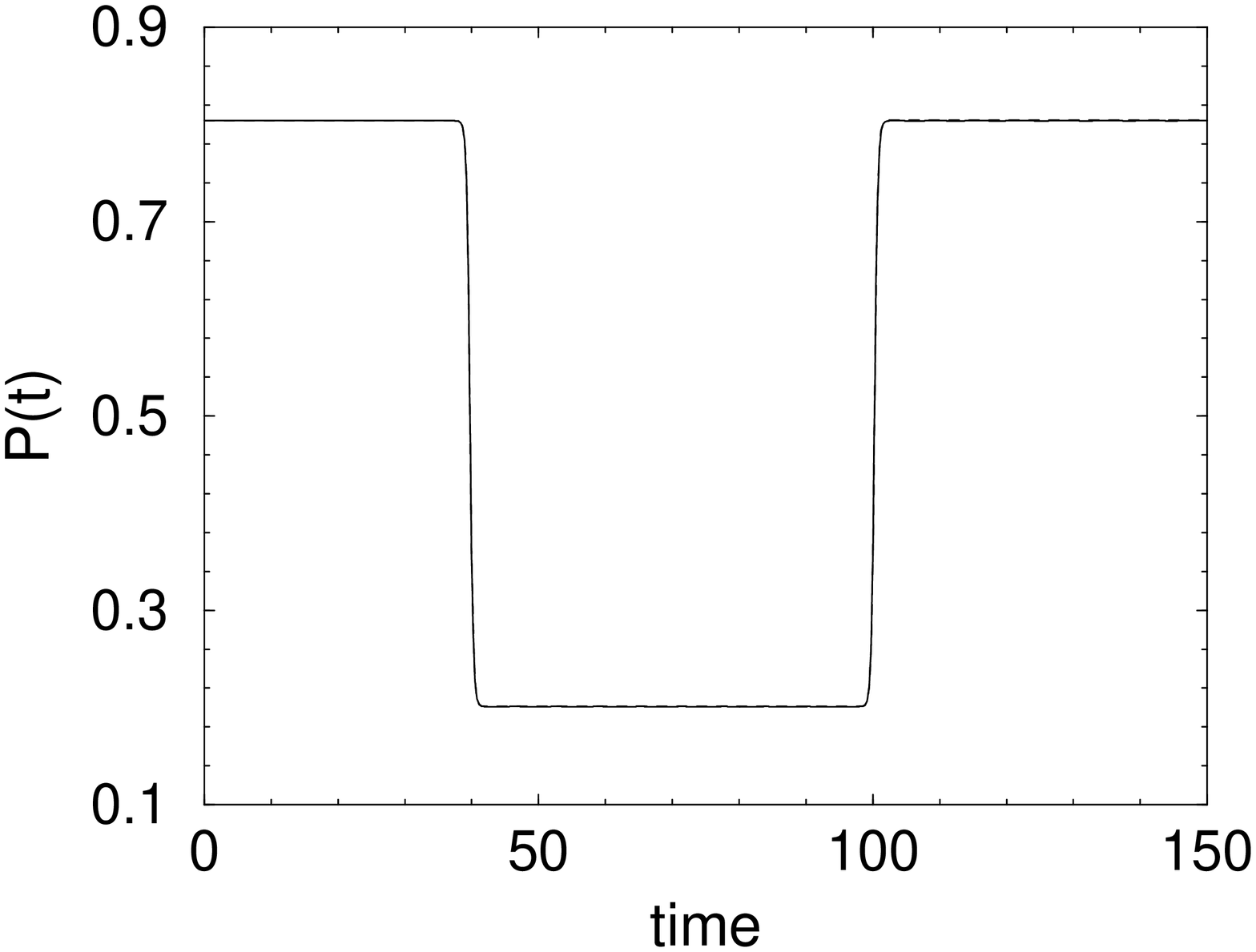}
\end{tabular}
\end{center}
\caption{Sine-Gordon kink dynamics in a \textit{fast} expanding
and later contracting space. Top-left panel: scale factor $ a(t)
$, Eq.~\eqref{adep} with $\Delta a = 1 $, $ t_0 = 40 $, $ t_1 =
100 $ and $ \Delta t = 0.5 $. Top-right panel: spatial profile of
the kink $ \phi(x,t_{fix}) $ at times $t_{fix}=37.5$ (solid line),
$t_{fix}=75$ (long-dashed line) and $t_{fix}=150$ (dashed line).
Three following panels: position of the center of the kink $
X_{phys}(t) $, peculiar kink velocity $ V_{pec}(t) $, and width of
the kink $ l_{phys}(t) $ as functions of time given by the numerical 
simulations of the PDE (solid line), the numerical solutions of the 
CCs equation 
(long-dashed line), and the
adiabatic approximation (dashed line). Bottom-right panel:
momentum of the kink $ P(t) $ exact solution. Initial conditions:
a kink with $ X(0) = 10 $ and $ \dot X(0) = 0.1 $. }
\label{figFastExpCon}
\end{figure}

When the expansion/contraction is faster the adiabatic
approximation breaks, and non-adiabatic effects appear as
radiation (compare the kink profiles in Fig.~\ref{figSlowExpCon}
and Fig.~\ref{figFastExpCon}) and as a change in the final width
and velocity of the kink [that are no longer those predicted by
Eqs.~\eqref{ladia}-\eqref{vadia}]. See Fig.~\ref{figFastExpCon},
that corresponds to the non-adiabatic regime $ H \sim \Delta a /
\Delta t = 2 $ and $ \dot H \sim \Delta a / (\Delta t)^2 = 4 $.
The CCs evolution equations predict strong oscillations in the
velocity and the width of the kink. In the numerical integration of
the complete evolution equations [Eq.\ (\ref{KGevoleq})] the
\emph{oscillations} are indeed present in the width. However, there are
other degrees of freedom to which the energy can be transferred.
This results in the
\emph{damping of the oscillations through radiation} emission. The
loss of energy can also be easily shown noting that after an
expansion-contraction cycle the kink moves slower (see
Fig.~\ref{figFastExpCon}).

We would like to stress that all the previous comments about slow and fast
expansions/contractions apply both for the sine-Gordon and the
$\phi^4$ potential. We have performed numerical simulations and
comparisons of the adiabatic approximation, the CCs evolution
equations, and the full evolution equation for both potentials
obtaining analogous results.

\bigskip


\section{Conclusions}\label{sConclusions}

We have studied the effects on a kink of slow and fast
expansions/contractions of the media, both for the sine-Gordon 
and for the $ \phi^4 $ equations.
We have used the Rice {\it Ansatz} \cite{rice} in the
continuity equations for the momentum and the energy, in order
to obtain the evolution equations for the width and the center
of a kink in an expanding/contracting space. In general this
set of ODEs cannot be solved analytically and the solution must be
obtained numerically. However, this effective description in terms
of the collective coordinates usually gives relevant information
and insight of the evolution and the phenomena involved
\cite{prl,KM2,scott,siam,Salerno}. In our case the CCs approximation is able
to describe very accurately the dynamics of the kink when the process is slow,
even when the variations of $ a $ are large.
We have found that the main effects of a \emph{slow}
expansion/contraction are the change of the width and the velocity 
of the kink, following the \emph{adiabatic
relations} \eqref{ladia} and \eqref{vadia}. In addition, the
collective coordinate evolution equations for the width and the
center of the kink are able to take into account the small deviations from
adiabaticity (excluding explicitly the radiative effects, because they 
are related to the transference of energy to other degrees of freedom).

An important consequence of the adiabatic relations is that the
speed of a kink can be tuned by slowly expansion/contraction of 
space. Thus, this result provides a mechanism to \emph{control the
speed of a kink} whenever we can expand and contract the media
where it propagates.

On the other hand, \emph{fast} expansions/contractions break the
adiabatic approximation, giving rise to \emph{non-adiabatic
effects}, as for example radiation, that imply a change in the final width and
velocity of the kink (i.e., they are no longer those predicted by
the adiabatic relations). The collective coordinate evolution
equations predict strong oscillations in the width of the kink.
In the full problem, these oscillations are also present. However,
there are other degrees of freedom to which the energy of these
\emph{oscillations} can be transferred. This results in emission of
\emph{radiation} and in the \emph{damping} of these oscillations.
In this sense, the fast expansion/contraction of space
provides a method to implement a
fast change in the speed of the kink, and also to generate phonons.

\appendix
\section{Expanding spaces}\label{sTechnical}
This appendix includes some technical details related to 
expanding spaces.

\subsection{Metric}

The metric in an homogeneous and isotropous flat space is given by
\be
ds^2 = dt^2 - a^2(t) dx^2 = g_{\mu\nu} dx^\mu dx^\nu \;,
\ee
where $ a(t) $ is the scale factor for the expansion/contraction,
the metric tensor is
\be
g_{\mu\nu} = \left( \ba{cc} 1 & 0 \\ 0 & -a^2(t)  \ea \right) \;.
\ee
Its inverse, $ g^{\mu\nu} $, defined by $ g_{\mu\nu}
g^{\nu\lambda} = g_{\mu}^{\;\;\lambda} =
\delta_{\mu}^{\;\;\lambda} $ (with $ \delta_{\mu}^{\;\;\lambda} $
a Kronecker delta) is
\be
g^{\mu\nu} = \left( \ba{cc} 1 & 0 \\ 0 & -\displaystyle{\frac1{a^2(t)}}  \ea
\right)
\ee
The contraction with $ g_{\mu\nu} $ lowers indices, and the
contraction with $ g^{\mu\nu} $ raises indices.

\subsection{Invariant element of space-time volume}

Defining
\be
g \equiv -\mbox{Det} g_{\mu\nu} \;,
\ee
it can be shown (\cite{weinberg} pages 98-99) that
\be
\sqrt{g} dtdx
\ee
is an invariant volume element under general coordinate
transformations. ($ dtdx $ transforms with the Jacobian, while $
\sqrt{g} $ transforms with the inverse of the Jacobian.)

Therefore the appropriate relation between the action $ S $, and
the Lagrangian density $ {\cal L} $ is
\be
S = \int{dt dx \sqrt{g} {\cal L} } \;.
\ee
This definition implies that if the action is invariant under
general coordinate transformations the Lagrangian density is also
invariant. In our case $ \sqrt{g} = a(t) $.

\subsection{Energy-momentum tensor}

The energy-momentum tensor is
\be
T^{\mu\nu} = \partial^\mu \phi \partial^\nu \phi - g^{\mu\nu}
{\cal L} \;.
\ee
Some useful relations to calculate its components are
\bea
 \partial^x \phi &=& g^{x\sigma} \partial_\sigma \phi = g^{xx}
\partial_x \phi = - \frac{\phi_x}{a^2} \;, \\
 \partial^t \phi &=& \partial_t \phi = \phi_t \;.
\eea
The time-time component of the energy-momentum tensor or energy
density is
\bea
\rho_E \equiv T^{tt} &=& \partial^t \phi \partial^t \phi - g^{tt}
\left[ \frac12 \phi_t^2 - \frac12 \frac{\phi_x^2}{a^2} - U(\phi)
\right] =  \nonumber \\
&=& \frac12 \phi_t^2 + \frac12 \frac{\phi_x^2}{a^2} +
U(\phi) \;,
\eea
the space-time component or momentum density is
\be
 \rho_P \equiv T^{xt} = T^{tx} = \partial^t \phi \partial^x \phi - g^{tx}
{\cal L} = -\frac{1}{a^2} \phi_t \phi_x \;,
\ee
and the space-space component or momentum current is
\be
 j_P \equiv T^{xx} = \frac{1}{a^2} \left( \frac12 \phi_t^2 +
 \frac12 \frac{\phi_x^2}{a^2} - U(\phi) \right) \;.
\ee

\section{Equivalence among different CC approaches}\label{equivalencia}

In this appendix we show that the evolution equations for the collective
coordinates obtained with the derivation of the momentum and the energy
\cite{emderivation} (that we used in Section III) are the same as those
obtained with the Lagrangian method \cite{lagform,saan} and with the
so-called {\it GTWA} \cite{gtwa,prl} (based on projection techniques).
In all cases, we use  the Rice {\it Ansatz} as an approximated kink-like
solution of the following perturbed nonlinear Klein-Gordon equation
\be \label{KGdis}
\phi_{tt}-\frac{\phi_{xx}}{a^2(t)} = -\frac{dU}{d\phi} -
[H(t)+\beta(t)] \phi_t + z(x,t,\phi)\;,
\ee
where $ \beta(t) $ is a time dependent damping coefficient and $
z(x,t,\phi) $ represents a generic perturbation on the system. Note that
for $\beta=0$ and $z(x,t,\phi)=0$ we recover the system
(\ref{KGevoleq}), introduced in Section II.

\subsection{Generalized Lagrangian Formalism}

Introducing a new time variable
\be
\tau \equiv \int_0^t{\frac{dt'}{c(t')}}\;;
\ee
where $c(t)$ is given by
\bea
c(t) &\equiv& \exp\left\{\int_0^{t}{[H(t') + \beta(t')]dt'}\right\}
\nonumber
\\&=& \frac{a(t)}{a(0)} \exp\left\{\int_0^t{\beta(t') dt'}\right\} \;,
\eea
Eq. \eqref{KGdis} becomes the dissipationless equation,
\be \label{KGnodiss}
\frac{\phi_{\tau\tau}}{C^2(\tau)} - \frac{\phi_{xx}}{A^2(\tau)} =
- \frac{dU}{d\phi} + Z(x,\tau,\phi) \;,
\ee
with $ C(\tau) = c(t(\tau)) $, $ A(\tau) = a(t(\tau)) $, and $
Z(x,\tau,\phi) = z(x,t(\tau),\phi) $.

The evolution equation \eqref{KGnodiss} can be obtained from the
action,
\be \label{genact}
S = \int{ d\tau dx {\cal L}} = \int{d\tau dx \left\{ \frac12
\phi_\tau^2 - \frac12 \frac{C^2(\tau)}{A^2(\tau)} \phi_x^2 -
C^2(\tau) [U(\phi) + M(x,\tau,\phi,\phi_x)] \right\}}
\ee
$ {\cal L} $ being the Lagrangian density, and $
M(x,\tau,\phi,\phi_x) $ an ``Euler-Lagrange integral'' of $
Z(x,\tau,\phi) $. Some of the possible integrations are
\begin{itemize}
\item the $ \phi $ functional integral,
\be \label{funcional}
M(x,\tau,\phi) = - \int{ {\cal D}\phi \; Z(x,\tau,\phi)}\;,
\ee
\item and the following integration for the particular case where $
Z(x,\tau,\phi) = F(x,\tau)G(\phi) $ \cite{saan},
\be \label{cuenda}
M(x,\tau,\phi,\phi_x) = \phi_{x}\, G(\phi)\, \int_{x_{0}}^{x}{ dx'
\; F(x',\tau)}\;.
\ee
\end{itemize}

Once we define the Lagrangian,
\be
L = \int_{\infty}^{-\infty} dx {\cal L} \;,
\ee
the Euler-Lagrange evolution equations for the collective variables
are obtained inserting a given \emph{Ansatz} in this
Lagrangian and using either the expression (\ref{funcional}) or
(\ref{cuenda}). This procedure is the generalization of the Legrangian 
formalism developed in \cite{lagform}.

Using the Rice Ansatz defined by (\ref{rice-sg}) and (\ref{rice-phi4})
(for the sG and $\phi^4$ potentials, respectively) and after some
straightforward calculations, we obtain the Lagrangian
as a function of our collectives variables and theirs derivatives
with respect to $\tau$
\bea \label{lag2}
L(X,X',l,l')&=&\frac{M_0 l_0}{2 l} \left (X' \right )^2+
\frac{\alpha M_0 l_0}{2 l} \left (l' \right )^2-\\ \nonumber
&&\frac{1}{2} M_0 C^2(\tau) \left
(\frac{l}{l_0}+\frac{l_0}{A^2(\tau) l} \right)- C^2(\tau)
\int_{-\infty}^{+\infty}dx \,M(x,\tau,\phi,\phi_x). 
\eea
Replacing $L(X,X',l,l')$ into the Euler-Lagrange equations,
\begin{eqnarray}
\frac{d}{d\tau} \left (\frac{\partial L}{\partial X'} \right) &=&
\frac{\partial L}{\partial X}, \\
\frac{d}{d\tau} \left (\frac{\partial L}{\partial l'} \right) &=&
\frac{\partial L}{\partial l};
\end{eqnarray}
and rewriting the equations of motion in the time variable $t$, we
get
\begin{eqnarray} \label{ccex2}
&& \dot{X} = \frac{a(t) P(t) l(t)}{M_{0} l_{0}}, \\ \label{ccep2}
&& \frac{dP}{dt} =- \left (\beta+2H \right) P-\frac{1}{a(t)}\left \{\int_{-\infty}^{+\infty}d\theta z(X+\theta l,t,\phi) \phi_{\theta}\right \} , \\
\label{ccel2}
&& \alpha \left [ \dot{l}^2-2 \left (\beta +H \right
)l\dot{l} -2l\ddot{l}\right] = \frac{l^2}{l^2_0} \left
(1+a^2(t)\frac{P^2}{ M_0^2} \right )-\frac{1}{a^2}
\nonumber \\
&\,& \qquad \qquad \qquad \qquad \qquad \qquad
+\frac{2 l^2}{M_0 l_0}
\int_{-\infty}^{+\infty}d\theta z(X+\theta l,t,\phi) \theta \phi_{\theta},
\end{eqnarray}
where $\theta=[x-X(t)]/l$ and the dots denote the derivative with
respect to $t$. These are the evolution equations for the
collective variables obtained for either of the two previous 
expressions of $ M(x,t,\phi,\phi_x) $ Eq.~\eqref{funcional} or
Eq.~\eqref{cuenda} (in their respective regime of validity).
Taking in these equations $\beta=0$ and $z=0$, we obtain
the Eqs.\ (\ref{ccex}), (\ref{ccep}) and (\ref{ccel})
derived in Section III.

\subsection{GTWA}

In order to apply the {\it GTWA} we rewrite Eq.~(\ref{KGdis}) as
\bea \label{GTWAeq}
\dot{\phi} &=& \psi, \nonumber \\
\dot{\psi}  &=& \frac{\phi_{xx}}{a^2(t)} -\frac{dU}{d\phi} -
[H(t)+\beta(t)] \psi + z(x,t,\phi).
\eea
The procedure to obtain the CCs equations with the GTWA consists
of inserting our specific functional form for $\phi$,
Eq.~(\ref{rice-sg}) for sine-Gordon 
or (\ref{rice-phi4}) for $\phi^4$ potential, into Eq.~(\ref{GTWAeq}),
multiplying the first equation by ${\partial \psi}/{\partial X}$
and the second one by ${\partial \phi}/{\partial X}$, taking their
difference and integrating over $x$, equating the result with
zero; and repeating the same procedure with ${\partial
\psi}/{\partial l}$ and ${\partial \phi}/{\partial l}$. This
gives to the same ordinary differential equations for $X(t)$,
$P(t)$, and $l(t)$ as those obtained in the previous subsection [see
Eqs.\ (\ref{ccex2})-(\ref{ccel2})].

\acknowledgments

We  acknowledge financial support from  the Ministerio de Ciencia
y Tecnolog\'{\i}a of Spain under grants No. BFM2003-02547/FISI
(FJC), No. NAN2004-09087-C0303 (FJC), and No. FIS2005-973 (EZS,NRQ), and
from the Junta de Andaluc\'{\i}a through the projects FQM-0207 and 00481 
(EZS,NRQ). In addition, EZS thanks the University of Sevilla for
its financial support.

\end{document}